# White Paper

# "Crowdsourced Network and QoE Measurements – Definitions, Use Cases and Challenges"

(Version 26.03.2020)

## List of Editors

Any comments or suggestions should be sent to the editors:

- Tobias Hoßfeld, University of Würzburg, Germany, tobias.hossfeld@uni-wuerzburg.de
- Stefan Wunderer, Nokia Solutions and Networks, stefan.wunderer@nokia.com

## List of Authors

The chapters were led by different chapter leads.

- Fabrice Guillemin, Chapter 1: Definition
- Andrew Hall, Chapter 2: Use Cases
- Michael Seufert, Chapter 3: Challenges

The complete list of authors is provided in alphabetical order.

- André Beyer
- Andrew Hall
- Anika Schwind
- Christian Gassner
- Fabrice Guillemin
- Florian Wamser
- Krzysztof Wascinski
- Matthias Hirth
- Michael Seufert
- Pedro Casas
- Phuoc Tran-Gia
- Stefan Wunderer
- Tobias Hoßfeld
- Werner Robitza
- Wojciech Wascinski
- Zied Ben Houidi

## Please cite this white paper as

"White Paper on Crowdsourced Network and QoE Measurements – Definitions, Use Cases and Challenges (2020). Tobias Hoßfeld and Stefan Wunderer, eds., Würzburg, Germany, March 2020. doi: 10.25972/OPUS-20232."

# White Paper

# "Crowdsourced Network and QoE Measurements – Definitions, Use Cases and Challenges"

## Preface

This white paper is the outcome of the Würzburg seminar on "Crowdsourced Network and QoE Measurements" which took place from 25-26 September 2019 in Würzburg, Germany. International experts were invited from industry and academia. They are well known in their communities, having different backgrounds in crowdsourcing, mobile networks, network measurements, network performance, Quality of Service (QoS), and Quality of Experience (QoE). The discussions in the seminar focused on how crowdsourcing will support vendors, operators, and regulators to determine the Quality of Experience in new 5G networks that enable various new applications and network architectures. As a result of the discussions, the need for a white paper manifested, with the goal of providing a scientific discussion of the terms "crowdsourced network measurements" and "crowdsourced QoE measurements", describing relevant use cases for such crowdsourced data, and its underlying challenges. During the seminar, those main topics were identified, intensively discussed in break-out groups, and brought back into the plenum several times. The outcome of the seminar is this white paper at hand which is – to our knowledge – the first one covering the topic of crowdsourced network and QoE measurements.

## Introduction

The term crowdsourcing was first used by Jeff Howe and Mark Robinson in 2005, to describe how businesses were using the Internet to outsource work to the crowd. Interestingly enough, Howe and Robinson were journalists, not scientists or engineers.

Principally, crowdsourcing is an old concept. In 1849, for instance, some 150 volunteer weather observers all over America were set up by Joseph Henry, the Smithsonian Institution's first secretary. They used the telegraph to make new weather information available to the public on a daily basis. For instance, volunteers tracked a tornado passing through Wisconsin and sent the findings via telegraph to the Smithsonian. Henry's project was very successful and is considered the origin of the National Weather Service. The weather project is emblematic as it contains the two essentials for crowdsourcing: volunteers and a network.

Crowdsourcing is currently gaining a lot of attention in the telecommunication industry and research, especially in the context of mobile networks. Measurements from end user perspective are essential to detect or to understand upcoming problems in networks, and therefore indispensable for improving QoS and enhancing QoE. Using the end user devices for gaining crowdsourced measurements and with subjective QoE studies on the user level, we can gain a much better holistic understanding of the impact of network challenges or issues on the quality experienced by end users.

Crowdsourced measurements offer interesting opportunities, and they can be applied to various use cases like benchmarking of network operators, vendors, technologies or countries as well as for monitoring, planning and optimizing the network. In these cases, crowdsourcing allows getting insights beyond the network layer, i.e., on the application layer and user level. This makes crowdsourced measurements very valuable, and it extends current practices of operators on how to gain detailed information about their networks. The ultimate goal is to use data gained via crowdsourcing – combined with other network and user data – to improve QoE, but also for regulatory insights to gain e.g. coverage, network, and quality information.

However, crowdsourcing sometimes raises scepticism in the network performance community. Top managers start to rely on it or to draw significant comparisons (as mentioned above). Performance and optimization engineers are concentrating on crowdsourcing reliability, measurement failures, and comparisons to their deeply known traditional network data. When, e.g., a timeline of the throughput measurements shows a ditch due to a server change, they doubt the whole crowdsourcing method very quickly. Scientific methods and conclusions concerning crowdsourcing are therefore essential to provide a profound basis for the network performance community. This white paper paves the way for such fundamentals.

The goal of the white paper at hand is as follows. The definitions of the terms build a framework for discussions around the hype topic 'crowdsourcing'. This serves as a basis for differentiation and a consistent view from different perspectives on crowdsourced network measurements, with the goal to provide a commonly accepted definition in the community. The focus is on the context of mobile and fixed network operators, but also on measurements of different layers (network, application, user layer). In addition, the white paper shows the value of crowdsourcing for selected use cases, e.g., to improve QoE or regulatory issues. Finally, the major challenges and issues for researchers and practitioners are highlighted.

This white paper is structured as follows. In the following chapter, **definitions** of the terms crowdsourced network and QoE measurements are provided. As a matter of fact, the human crowd is an essential part. These definitions lead to various dimensions for a taxonomy of crowdsourcing, which includes the collection of subjective or objective data, incentives or allurement for the crowd, as well as the engagement or effort to execute the measurements. To this end, examples are discussed to validate our crowdsourced network measurement definitions. Along those identified dimensions, various use cases of crowdsourcing are analysed and classified in a structured way in the **use cases chapter**. Those use cases drive the test design and methodology. However, this will reveal particular challenges to be discussed in the **challenges chapter**. The challenges revolve around issues of validity and reliability and also include social and ethical aspects: the human crowd – which means people – are used to gain data, potentially without their explicit knowledge (Martin 2017). Even when their full anonymity is assured, it is possible that someone is making money or creating business value without revenue or obvious benefit to the end-user who is delivering the data.

# Chapter 1: Definition

## 1.1 Background

Crowdsourcing is a well-established concept in the scientific community and has enabled a huge number of new engineering rules and commercial applications. This section focuses on the current state of the art in crowdsourcing research and lays the foundation for our definition of crowdsourcing in the context of network measurements.

### 1.1.1 Crowdsourcing

The word *crowdsourcing* itself is a mix of the *crowd* and traditional *outsourcing* work-commissioning model. It was first introduced by Jeff Howe in 2005 (Howe, 2006). Since the publication, the research community has been struggling to find a definition of the term crowdsourcing (Estellés-Arolas, 2012; Kietzmann 2017; *ITU-T P.912,* 2016) that fits the variety of its applications and new developments.

For example, in ITU-T P.912, crowdsourcing has been defined as:

> ***Crowdsourcing*** *is obtaining the needed service by a large group of people, most probably an on-line community.*

The above definition has been written with the main purpose of collecting subjective feedback from users. For the purpose of this whitepaper, it is required to clarify this definition and thus, in general, the term crowdsourcing will be defined as follows:

> ***Crowdsourcing*** *is an action by an initiator who outsources tasks to a crowd of participants to achieve a certain goal.*

The following terms are further defined to clarify the above definition:

> *A **crowdsourcing action** is part of a campaign that includes processes such as campaign design and methodology definition, data capturing and storage, and data analysis.*
>
> *The **initiator** of a crowdsourcing action can be a company, an agency (e.g., a regulator), a research institute or an individual.*
>
> *Crowdsourcing **participants** (also "workers" or "users") work on the tasks set up by the initiator. They are third parties with respect to the initiator, and they must be human.*
>
> *The **goal** of a crowdsourcing action is its main purpose from the initiator's perspective.*

The outcome of a crowdsourcing **action** is the **crowd data**. The format of the **crowd data** is specified by the **initiator** and depends on the type of crowdsourcing **action**. For instance, crowd data can be results of computation in the case of a large scale computation experiments, analytics, measurement data, etc. In addition, the semantic interpretation of crowd data is under the responsibility of the **initiator**. The participants cannot interpret the **crowd data**, which must be thoroughly processed by the **initiator** to reach the **objective** of the crowdsourcing **action**.

Goals can be manifold and may include, for example:

- Gathering subjective feedback from users about an application (e.g., ranks expressing the experience of users when using an application)
- Leveraging existing capacities (e.g., storage, computing, etc.) offered by companies or individual users to perform some task
- Leveraging cognitive efforts of humans for problem solving in a scientific context

In general, an initiator adopts a crowdsourcing approach because of a lack of resources (e.g., running a large-scale computation by using the resources of a large number of users to overcome its own limitations) or to broaden a test basis much further than classical opinion polls. Crowdsourcing thus covers a large range of actions with various degrees of involvement by the participants.

In crowdsourcing, there are various methods of identifying, selecting, receiving, and paying users contributing to a crowdsourcing initiative and related services. Individuals or organizations obtain goods and/or services in many different ways from a large, relatively open and often rapidly-evolving group of crowdsourcing participants (also called users)**.**

The use of goods or information obtained by crowdsourcing to achieve a cumulative result can also depend on the type of task, the collected goods or information and final goal of the crowdsourcing task.

### 1.1.2 Roles and actors

Besides the goal, the other important aspect of a crowdsourcing action is the involved actors, namely the initiator and the participants.

The role of the initiator is to design and initiate the crowdsourcing action, distribute the required resources to the participants (e.g., a piece of software or the task instructions, assign tasks to the participants or start an open call to a larger group), and finally to collect, process and evaluate the results of the crowdsourcing action.

The role of the participants depends on their degree of contribution or involvement. In general, their role is described as follows. At least, offer their resources to the initiator, e.g., time, ideas, or computation resources. In tasks that require higher levels of contributions, participants might run or perform the tasks assigned by the initiator, and (optionally) report the results to the initiator.

Finally, the relationships between the initiator and the participants are governed by policies specifying the contextual aspects of the crowdsourcing action such as security and confidentiality, and any interest or business aspects specifying how the participants are remunerated, rewarded, or incentivized for their participation in the crowdsourcing action.

## 1.2 Crowdsourcing in the Context of Network Measurements

The above model considers crowdsourcing at large. In this section, we analyse crowdsourcing for network measurements, which creates **crowd data**. This exemplifies the broader definitions introduced above, even if the scope is more restricted but with strong contextual aspects like security and confidentiality rules.

### 1.2.1 Definition: Crowdsourced Network Measurements

Crowdsourcing enables a distributed and scalable approach to performing network measurements. It can reach a large number of end users all over the world. This clearly surpasses the traditional measurement campaigns launched by network operators or regulatory agencies able to reach only a limited sample of users. Primarily, crowd data may be used for the purpose of evaluating QoS, that is, network performance measurements. Crowdsourcing may however also be relevant for evaluating QoE, as it may involve asking users for their experience – depending on the type of campaign.

With regard to the previous section and the special aspects of network measurements, crowdsourced network measurements / crowd data are defined as follows, based on the previous, general definition of crowdsourcing shown above:

> ***Crowdsourced network measurements*** *are actions by an initiator who outsources tasks to a crowd of participants to achieve the goal of gathering network measurement-related crowd data.*
>
> ***Crowd data*** *is the data that is generated in the context of crowdsourced network measurement actions.*

The format of the crowd data is specified by the initiator and depends on the type of crowdsourcing action. For instance, crowd data can be results of computation in the case of a large scale computation experiments, analytics, measurement data, etc. In addition, the semantic interpretation of crowd data is under the responsibility of the initiator. The participants cannot interpret the crowd data, which must be thoroughly processed by the initiator to reach the objective of the crowdsourcing action.

An important aspect to consider is the contribution of human participants. In this paper, the case of distributed measurement actions solely made by robots, IoT devices or automated probes is therefore excluded.

Additionally, we only consider cases in which the participants consent to participate in the crowdsourcing action. However, this consent might vary from actively fulfilling dedicated task instructions provided by the initiator of the crowdsourcing action to merely accepting terms of services that include the option to analyse usage artefacts generated while interacting with a service.

It follows that in the present document, it is assumed that measurements via crowdsourcing (namely, crowd data) are performed by human participants aware of the fact that they are participating in a crowdsourcing campaign. Once clearly stated, more details need to be provided about the slightly adapted roles of the actors and their relationships in a crowdsourcing initiative in the context of network measurements.

### 1.2.2 Active and passive measurements

For a better classification of crowdsourced network measurements, it is important to differentiate between **active** and **passive** measurements. Similar to the current working definition within the ITU-T Study Group 12 work item "E.CrowdESFB" (*Crowdsourcing Approach for the assessment of end-to-end QoS in Fixed Broadband and Mobile Networks*), the following definitions are made:

> ***Active measurements*** *create artificial traffic to generate crowd data.*
>
> ***Passive measurements*** *do not create artificial traffic, but measure crowd data that is generated by the participant.*

For example, a typical case of an active measurement is a speed test that generates artificial traffic against a test server in order to estimate bandwidth. A passive measurement instead may be realized by fetching cellular information from a mobile device, which has been collected without additional data generation.

### 1.2.3 Roles of the actors

Participants have to commit to the participation in the crowdsourcing measurements. The level of contribution can vary depending on the corresponding effort or level of engagement, going from the simplest action of subscribing to or installing a specific application which collects data through measurements as part of its functioning – often in the background and not as part of the core functionality provided to the user –, to a more complex task-driven engagement requiring a more important cognitive effort, such as providing subjective feedback on the performance or quality of certain Internet services.

Hence, one must differentiate between **participant-initiated** measurements and **automated** measurements:

> ***Participant-initiated measurements*** *require the participant to initiate the measurement. The measurement data are typically provided to the participant.*
>
> ***Automated measurements*** *can be performed without the need for the participant to initiate them. They are typically performed in the background.*

A participant can thus be a **user** or a **worker**. The distinction depends on the main focus of the person doing the contribution and his/her engagement:

> *A **crowdsourcing user** is providing crowd data as the side effect of another activity, in the context of passive, automated measurements.*
>
> *A **crowdsourcing worker** is providing crowd data as a consequence of his/her engagement when performing specific tasks, in the context of active, participant-initiated measurements.*

The term "users" should therefore be used when the crowdsourced activity is not the main focus of engagement, but comes as a side effect of another activity – for example, when using a web browsing application which collects measurements in the background, which is a passive, automated measurement.

"Workers" are present when the crowdsourced activity is the main driver of engagement, for example, when the worker is paid to perform specific tasks, and is performing an active, participant-initiated measurement. Note that in some cases, workers can also be incentivized to provide passive measurement data (e.g. with applications collecting data in the background if not actively used).

In general, workers are paid on the basis of clear guidelines for their specific crowdsourcing activity, whereas users provide their contribution on the basis of a more ambiguous, indirect engagement, such as via the utilization of a particular service provided by the beneficiary of the crowdsourcing results, or a third-party crowd provider. Regardless of the participants' level of engagement, the data resulting from the crowdsourcing measurement action is reported back to the initiator.

The initiator of the crowdsourcing measurement action often has to design a crowdsourcing measurement campaign, recruit the participants (selectively or openly), provide them with the necessary means (e.g. infrastructure and/or software) to run their action, provide the required (backend) infrastructure and software tools to the participants to run the action, collect, process and analyse the information, and possibly publish the results.

The crowdsourcing measurement action should be in conformance with security and confidentiality policies, in particular the General Data Protection Regulation (GDPR) in Europe. The results (namely, measurements) of a crowdsourcing measurement action (crowd data) thus have to conform to the security and confidentiality rules of the action. This crowd data can be processed and interpreted only by the initiator. It is however crucial that participants are conscious of the purpose of the data collection, specified in the privacy terms of the used measurement software or campaign. From a business or commercial perspective, the initiator can either remunerate the participants or offer other incentives to participate in the crowdsourcing measurement action.

### 1.2.4 Dimensions of Crowdsourced Network Measurements

There are multiple dimensions to consider for crowdsourcing in the context of network measurements. A preliminary list of dimensions includes:

- Level of subjectivity (subjective vs. objective measurements) in the crowd data
- Level of engagement of the participant (participant-initiated or background) or their cognitive effort, and awareness (consciousness) of the measurement level of traffic generation (active vs. passive)
- Type and level of incentives (attractiveness/appeal, paid or unpaid)

Besides these key dimensions, there are other features which are relevant in characterizing a crowdsourced network measurement activity. These include scale, cost, and value; the type of data collected; the goal or the intention, i.e. the intention of the user (based on incentives) versus the intention of the crowdsourcing initiator of the resulting output.

In Figure 1, we have illustrated some dimensions of network measurements based on crowdsourcing. Only the subjectivity, engagement and incentives dimension are displayed, on an arbitrary scale. The objective of this figure is to show that an initiator has a wide range of combinations for crowdsourcing action. The success of a measurement action with regard to an objective (number of participants, relevance of the results, etc.) is multifactorial.

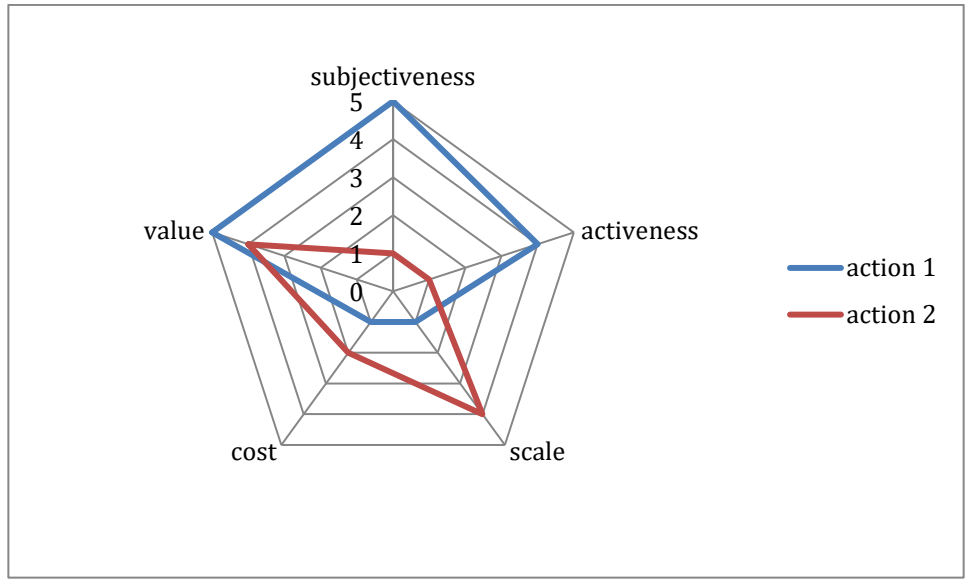

*Figure 1: Dimensions for Network Measurements Crowdsourcing definition, and Relevant Characterization Features (examples with two types of measurement actions)*

### 1.2.5 Example Crowdsourcing Approaches in Mobile Networks

Existing approaches to crowdsourcing in the mobile network industry are:

1. Active subjective testing, in which participants are recruited to actively record their perception of network quality. For example, a group of participants (in this case, workers) may be asked to view a video streamed over a network connection and rate the quality of the video on a scale between 1 and 5 (typically an input for mean opinion scoring). The incentive in this case is usually a financial reward.
2. Active objective testing, in which participants (workers) are provided with the means to record network performance via an application on their mobile device. The incentive in this case is not a financial reward, but access to the data that they collect. As a result, the initiator has less control over when tests are conducted. An example would be a speed-test application, where the user may initiate a test due to some circumstance (usually very good or very poor network conditions).
3. Passive objective testing, in which participants (users) passively provide data to the initiator as they go about their normal daily activities. This is usually achieved by embedding testing software into other applications which the participants install and use, informing them that data will be collected in this fashion as part of the terms and conditions of the application usage. As this testing is performed passively and under the control of the initiator, it provides the most representative set of measurements.

# Chapter 2: Use Cases

## 2.1 Objective

The goal of this section is to provide examples of use cases that can benefit from crowdsourced data, describing how it can be used, the extent to which it is relevant and the potential new value and insight it can bring over traditional methods.

## 2.2 Use case Overview

While there are many industries that can benefit from crowdsourcing approaches, the focus of this chapter will be those use cases most relevant to the mobile network industry. Table 1 summarises these use cases and relevant aspects of each, including the value, key beneficiaries and current methods of fulfilment.

When considering relevant mobile network use cases to discuss, it is helpful to consider the challenges faced by mobile network operators (MNOs). Differentiation between competing MNOs is largely based on two aspects: network coverage and network quality.

In general, network coverage is a function of site placement, site density and frequency bands used. Improving coverage can be achieved through **optimisation** of existing deployments or the deployment of a new frequency spectrum (**network planning**).

Network quality is often measured by performing certain activities on a live network and recording various key performance indicators (KPIs). For example, the downlink throughput capability can be measured by performing a file download of known size and monitoring the time taken to do so. KPIs typically include network availability, signal strength and quality, throughput (uplink and downlink), latency, loss, jitter and application-related statistics such as video buffering and quality.

The comparison between network quality measurements of different MNOs is termed **benchmarking**. It provides a competitive view of the market (which may be at a country or global level) and is a valuable piece of intelligence used in many departments of a network operator.

Benchmarking, optimisation and network planning are all important activities within a mobile network operator and are all use cases for crowdsourced data. In the next section, these and other use cases are discussed in more detail, with relevant focus on the impact crowdsourced data brings to each of them.

## 2.3 Example Mobile Network Operator Use cases

In order to ease understanding and comparison of different use cases, the following Table 1 contains a number of features for classifying the use cases. The following aspects are used to characterize a use case:

- **Purpose:** The goal of the use case, which sets the scene of test design and scope of methodology and measurements
- **Beneficiaries:** Departments and roles of those interested in the use cases and their measurements
- **Method of Fulfilment:** current and proposed methods of fulfilling the use case
- **Measurements:** the relevant types of data needed
- **Requirements:** every use case has its own degree of representative sample size, granularity, and geographic and temporal coverage

### 2.3.1 Table of Typical Use cases

*Table 1: Example Mobile Network Use cases*

| Use case | Purpose | Beneficiaries | Methods of Fulfilment | Measurements | Requirements |
|---|---|---|---|---|---|
| **Benchmarking** | Competitive positioning, market intelligence | Management and board reporting, strategic decision making | Drive/walk tests, crowdsourcing | Voice quality and availability, data throughput, latency, loss, jitter, signal strength, video quality, application measures | Mostly low granularity (from global/regional) with operator and technology split. Monthly or yearly time-scales. |
| **Planning** | Plan spectrum deployment, site selection, tune propagation models | Network planning teams, strategic planning | Geodata, drive/walk testing, crowdsourcing, operational support system (OSS) data | Signal level, signal quality, band usage, frequency usage | High data granularity (per sample). Regional context for a medium period of time |
| **Optimisation and Trouble-shooting** | Detect congestion and no-coverage areas; Detect and diagnose problems to improve the network | Radio engineering, network planning teams, customer support | Drive/walk tests, crowdsourcing, call tracing, OSS | Data throughput, signal strength, signal coverage, voice quality / availability | Highest possible data granularity (per sample). Local/focused context mainly (small area or city) for a short period of time |

## 2.3.2 Detailed Characterisation

### 2.3.2.1 Benchmarking

**Purpose and Value**

Benchmarking is an activity that provides an MNO and vendors with crucial intelligence regarding the relative performance of its own services and products as compared with the competition. This comparative analysis is used widely within an operator and often published to executive and C-level management. Understanding the position of an operators business in the market and the underlying reasons is powerful and influences decisions regarding the investment of both time and capital as well as feeding into optimisation, troubleshooting, and marketing activities.

**Methods of Fulfilment**

Benchmarking has typically been performed using data gathered from drive testing, which is the activity of deploying test equipment into vehicles and driving through an area of interest while the test equipment records measurements. Walk testing is similar but involves an individual carrying the test equipment into areas normally inaccessible by vehicles (such as shopping centres and airports).

Crowdsourcing is a new approach to providing the necessary data for benchmarking. In the terminology defined in chapter 1, the **initiator** of the crowdsourcing activity is sometimes the mobile network operator (in the case where the MNO develops and deploys their own crowdsourcing implementation) but can also be a third party crowdsourcing company that collects data through a **passive/active objective** approach and resells this data to the MNO.

**Crowdsourced Data Impact**

Drive testing and crowdsourced data largely complement one-another. While drive testing provides an enormous level of detail (offering 'layer-3' protocol and radio signalling information) as well as a high-level of control (in terms of which tests to conduct, which devices to use, where and when to test etc.), it can not provide a significant sample across both time and space due to logistical and cost reasons.

Conversely, crowdsourced data is able to provide significant sample sizes both temporally and spatially, across many devices, and can provide a number of service-level KPIs at a cheaper cost compared to drive test solutions, but is unable to offer the same level of low-level detail due to limitations imposed by handset operating systems, as well as regulatory restrictions such as the European General Data Protection Regulation (GDPR).

The emerging trend is that crowdsourced data is complementing existing drive testing activities when it comes to benchmarking. Crowdsourced data is often used at a national and regional level, with drive testing being used for smaller areas where more detail is needed. The coexistence and future integration of these two types of benchmarking approach will be an interesting development to observe over the coming years.

### 2.3.2.2 Planning

**Purpose and Value**

Network planning is largely concerned with the efficient deployment of new coverage and capacity, typically achieved through the deployment of additional spectrum or new cell site locations. When rolling out new spectrum, important decisions need to be made regarding optimal site locations, such as where coverage is already poor, non-existent or where competitors provide superior quality.

The ability to determine priority locations, predict network coverage and quality and develop cost-benefit analyses are all critical aspects of good network planning.

***Methods of Fulfilment***

Network planning is usually conducted using a team of highly specialised radio engineering teams and a selection of tools, the centrepiece usually a planning tool that can model radio propagation in conjunction with topographical maps, antenna configurations and clutter information (trees, buildings etc). These propagation models can be tuned for greater accuracy using real-world signal measurements sourced from drive testing or crowdsourced data.

In addition, executive-level decisions can impact network planning by steering effort towards certain marketing and sales goals, such as maximising the impact of initial 5G deployments by focusing on areas populated by early-adopters or high data-usage areas.

***Crowdsourced Data Impact***

Crowdsourcing data can play a role in deciding which sites to deploy first, by providing actual user experience KPIs in almost all locations and times of the day, highlighting specific conditions such as congestion, high data usage and poor end-to-end customer experience.

Additionally, network planning teams will be involved in the process of deactivating older technologies such as 3G. In this case, crowdsource data could identify areas where 4G capable devices in 4G coverage are connected to 3G, giving the operator valuable information about the potential impact of a switch-off or "sunset".

Crowdsourcing also provides valuable data regarding no-coverage areas, where the operator has no coverage available. Such areas are rarely discovered through drive testing and only detected with a larger base of samples only available through crowdsourced approaches.

### 2.3.2.3 Optimisation and Troubleshooting

***Purpose and Value***

As discussed in the introduction, coverage is a major concern for an MNO, as lack of coverage, or poor quality coverage, is an influential driver in causing the loss of subscribers (churn). In addition to providing good quality coverage, quality of experience and service are also crucial factors in affecting churn, as subscribers seek other competing networks if they are not able to attain the expected level of service they desire.

Optimisation is the process of analysing coverage and quality for areas of poor performance, establishing root causes (troubleshooting) and implementing changes that make improvements. Optimisation can also be triggered due to changes in network parametrization (e.g. tweaks to mobility thresholds) or the introduction of new radio features (e.g. MIMO or carrier aggregation).

Poor performing areas, low service quality and network faults can be discovered through customer complaints, benchmarking and internal alarms. However it is important to note that service quality can be impacted without the network necessarily being aware of a problem, for example in the case of no-service areas where subscribers are not registered to the network (invisible to the network operator).

Factors that contribute to poor performing areas include weak signal strength, excessive interference, overshooting cells and network misconfiguration (such as missing neighbour relations or crossed feeders). Due to human factors there is a geospatial and temporal aspect to each of these potential problems.

*Methods of Fulfilment*

Once a potential network issue has been identified, mobile network operators typically use a combination of tools to diagnose potential root-causes before deploying and testing potential solutions, including operational support systems (OSS), propagation models, drive test and crowdsourced data.

Service issues can stem from various factors, leading to the following as a list of necessary information to detect and diagnose such problems:

- Device models and capabilities
- Geographic location
- Radio access technology
- Cell information - such as cell identifiers, sectors, scrambling codes, frequencies
- Time of day
- Quality of experience KPIs such as latency, loss, throughput, video quality

*Crowdsourced Data Impact*

Crowdsourced data can provide a significant edge to an operator regarding the detection of network issues due to the sheer volume of data available that covers many of the requirements listed above, particularly regarding geographic, device and temporal diversity of samples which can be difficult to provide through other sources.

While crowdsourced data can provide a wealth of information that can be used to detect and, to some extent, diagnose network issues, it is not able to provide detailed radio signalling data that drive testing data can provide. As a result, an effective option is the combination of these two datasets, with crowdsourced data taking the role of detection and initial analysis (e.g. by indicating an issue with a new device firmware) and drive testing providing the detailed logging information that can be used to determine the underlying causes.

Furthermore, crowdsourced data can be used to verify changes made to the network to resolve problems by performing a before/after analysis, a much more cost-effective solution than rolling out a drive testing team to test the affected area.

# Chapter 3: Challenges

Given the relatively young age of the method of collecting data through crowdsourcing and its broad application area, there is a number of open challenges associated with it. In contrast to classical psychological research methods and fine-tuned laboratory measurement systems, on the one hand, the mass of the crowd provides researchers with vastly greater amounts of data. On the other hand, however, it also includes more unknown factors that cannot be measured, which may potentially influence the results and lead to unreliability or lack of trust in crowd data.

The new paradigm of designing crowd campaigns, incentivizing users, collecting the data, storing and processing it, and drawing conclusions from massive volumes of data, has its own unique characteristics that need to be addressed by managers, researchers and data scientists, particularly when it comes to the interpretation of the data and social/ethical aspects of the measurement campaign itself.

This chapter lists those challenges as they may be encountered in the context of a typical crowdsourcing study. We are not assuming a particular use case here – apart from the general scope of performing crowdsourced network measurements – and hence, the listed challenges should be applicable to most crowdsourcing campaigns within that scope.

As these challenges are described, frequent references to three overarching concepts will be made:

1) Validity
2) Reliability
3) Representativeness

These concepts are closely related but distinct. *Validity* refers to the "confidence that a given finding shows what it purports to show" (Haslam 2003). Often times, researchers risk making the wrong claims from data that was interpreted in the wrong way, or data that was measured in the wrong fashion, to begin with, or cause-effect relationships that have been falsely interpreted. A valid measurement is the ultimate goal of a crowdsourcing campaign, but oftentimes, validity is compromised when crowd users – or researchers – are biased, or when the wrong assumptions are made.

Next, one must contrast validity with *reliability*, which is the "confidence that a given empirical finding can be reproduced" (Haslam 2003). When aiming for reliable measurements, one wants to achieve the same results given the same context – be it known or unknown – to a certain extent, knowing, of course, that nothing can be measured perfectly accurately.

A good conceptual image showing the difference between validity and reliability is shown in Figure 2, which originates from ITU-T Rec. G.1011:

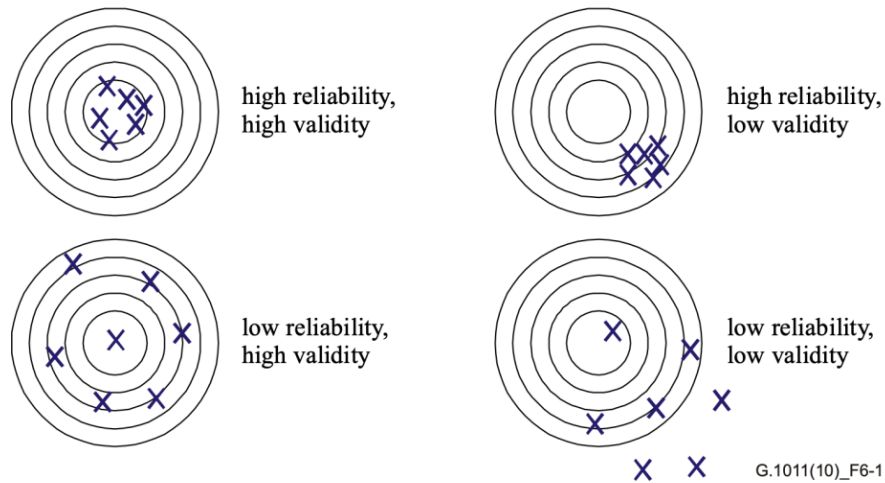

*Figure 2: Validity vs Reliability according to ITU-T Rec. G.1011*

Finally, *representativeness* is the degree to which the data sample is representative for the assumed population, or that an outcome is representative for the targeted population. As can be seen later, for example, reliability may strongly influence the (statistical) representativeness.

In the remainder of this chapter, these three concepts are instantiated – along with several others – during the typical campaign stages of 1) design and methodology, 2) data capturing and storage, and 3) data analysis. A roadmap of challenges during a crowdsourced measurement campaign can be seen in Figure 3. While it is not the purpose of this chapter to provide conclusive answers to all of the identified challenges, it should provide a list of topics that researchers may use for improving their crowdsourcing campaigns.

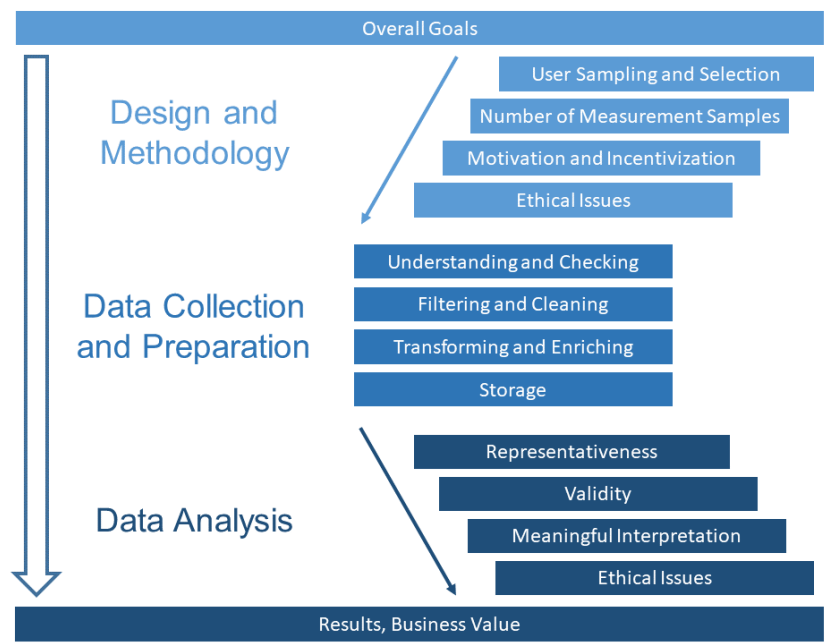

*Figure 3: Challenges during Crowdsourcing Campaigns*

## 3.1 Design and Methodology

To collect valid, reliable, and representative data, it is important to carefully design the measurement campaign before the actual measurements take place. Doing so, it is essential to keep in mind that already this early stage includes many challenges that can significantly affect later results.

The factors that come into play at the design stage have been grouped in this section.

### 3.1.1 Overall Goals

Even before designing the methodology, it is crucial to define the **general goals**: What exactly should be investigated? Which tools are needed for this purpose? Which methodology is most suitable and why? Which statements can be made later with the collected data and which not?

Here, not only the scope but also the limitations of the chosen methodology must be considered to adjust the design of measurements. In particular, the results of a measurement campaign may not be directly usable for other purposes, and hence the validity of statements made on the collected data can be criticized. For example, if a measurement campaign is designed to quantify download speeds, it is doubtful whether the data can be used to quantify the quality of telephony.

### 3.1.2 User Sampling and Selection

Once goals have been specified, the methodology of how to conduct the measurements must be put into focus. Here, the first question is how to design the measurements in order to collect representative results. In the crowdsourcing use case, this means: which users must be selected for the measurements to represent the target population?

Hence, the target population must be defined first; the influence factors must be determined to avoid sampling biases as illustrated in Figure 4. These biases occur, for example, when the users and their end devices are not sampled in a uniformly random way from the target population. For example, recruiting students for a measurement campaign might not be representative of the whole population of a city or country in terms of activity hours and Internet usage. That is the reason why the sampling method should be selected carefully considering the target population and the scope of the results.

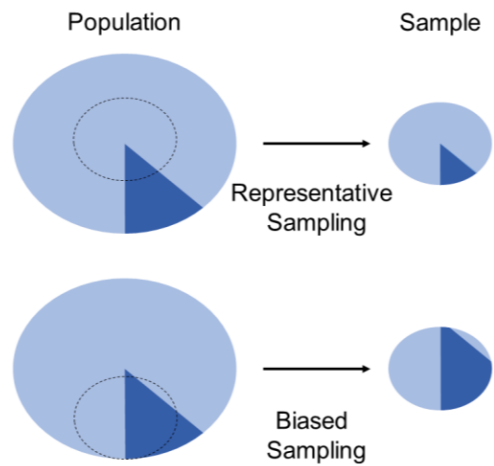

*Figure 4: Illustration of Sampling Bias*

### 3.1.3 Number of Measurement Samples

Regarding data reliability, a high number of users and measurements are essential. Usually, there are much more data samples available from passive, background measurements. One reason for this is that they have a much lower impact in terms of CPU and battery consumption compared to active measurements, and they also do not consume a significant data quota from the end user's tariff. They can be scheduled by the operator, and one does not have to wait for the user's participation.

Often, operators are interested in the end user's experience, where network speeds (throughput) and latency play a significant role. Here, active measurements provide the most reliable results, since realistic and valid application-level traffic can be generated, which is not always possible in the background due to device limitations.

As the active measurements cannot be collected very often, the number of samples could become low, depending on the selected time frame or location. The total amount of samples might be high on a national level, but very low on a city or street level. Hence, the data may be enough for marketing purposes and a trend analysis per country or mobile network, but not enough for network optimization – down to the street level. Depending on the data analysis requirements, the whole dataset may need to be further filtered based on time, MNO, radio technology, region, city, node/cell identifier, street level, or hot spot locations such as stadiums. Because of this, with an increasing application of filters, the number of collected samples can become too small. Thus, it has to be verified that after filtering enough data remains to assure reliable results, i.e., the number of samples is statistically relevant and sufficient to draw reliable conclusions.

### 3.1.4 Motivation and Incentivization

After selecting the measurement crowd – unless measurements are performed completely automatically without the user's involvement – users have to be motivated to participate in the measurement campaign. To do this, there are various methods, from monetary incentives to considerations of availability, such as the free use of an application or an obvious benefit from performing the measurements (e.g., gaining insights from the campaign results). Table 2 summarizes several advantages and disadvantages of the design choices.

Depending on the incentive, the influence of the chosen method on the overall validity of the results cannot be neglected. If someone is paid to participate, the results can be significantly different from the results of a user who participates for free – particularly when user opinions are involved. For example, a paid user is very likely to perform measurements when and where he or she is told to do them, but any subjective rating collected may be biased by the fact that he or she is paid to provide them. Hence, the human factor creates a high possibility that data is manipulated, for example by providing invalid subjective feedback on purpose. Since the time, location, and network condition of the device can be fully controlled by the user, this could lead to good or bad measurement results (e.g., high or low throughput values), which may have been provided on purpose, in order to manipulate the operator's rating in a positive or negative manner.

Also, a non-paid user cannot be controlled to perform measurements in the same regularity as a paid user because of external motivation factors such as dissatisfaction with the network or service. A voluntary measurement may, for example, be performed because someone is particularly unhappy with his or her connection, or vice-versa: the user wants to prove that his or her network can achieve great performance. This may skew the results.

While passively crowdsourced data collected in the background is less prone to manipulation, it can be more statistically random because the measurements are triggered without interaction from the end-user. The crowd data are automatically collected in different situations of an end user's daily life – including stationary periods and mobility. Such measurements are the main interest of operators who use the data for optimizing or benchmarking their network. A drawback of this strategy is that the amount of data used for background measurements (e.g., connection throughput) is much more limited in order to keep the consumed CPU, battery, and data volume low, and the impact on the end user's tariff budget on an acceptable level.

*Table 2: Advantages and Disadvantages of Design Choices between Paid/Unpaid Participants and Active/Passive measurements*

| Design choices | Advantages | Disadvantages |
|---|---|---|
| **Paid participants** | - Selection of participants<br>- More control over measurement situation and execution | - Payment of participants incurs costs<br>- Potentially biased subjective feedback<br>- Higher incentive for cheating |
| **Unpaid participants** | - No payment of participants → lower costs<br>- More intrinsically motivated participation | - No selection of participants<br>- Less control over measurement execution and situation |
| **Active, participant-initiated measurements** | - Richer possibilities for measurements<br>- Parallel activities less likely | - Triggered only by user interaction<br>- Prone to manipulation |
| **Passive, background measurements** | - Distribution of measurements over different situations<br>- Hard to manipulate | - Limited possibilities for measurements<br>- Potential influence of parallel activities |

### 3.1.5 Ethics and Legal Issues

Further challenges in recruiting users comprise ethical and legal issues. Although lots of data can be collected from end-users, the challenge is to decide how far to go. How much data is really needed to achieve the (previously defined) goal?

One imperative should be to not collect all available data only because it is technically doable and might prove useful later. Furthermore, one should be especially careful when collecting sensitive or personally identifying data such as exact geolocation, IP addresses, mobile device identifiers, or phone numbers. Here, it is important to only collect a data point if needed, and that sensitive data are anonymised immediately after use (e.g., when translating an IP address to an ISP). It should not be stored for longer periods than required, and the data should not reveal sensitive information about the end-user. Analysts must only be able to access data in a level of detail that is required to perform the necessary analysis steps.

In addition, before collecting the data, users should be informed about the amount and frequency of data collection. This is not just an ethical, but also a legal requirement in many jurisdictions. Here, questions arise whether it is enough to state this information in the terms and conditions, where only a few users will read it, or whether the data collection should be listed more prominently.

Particularly in the case of automatic background measurements, users may not be aware of the data collection. Depending on the legal framework of the country, in which the measurements are performed, explicit user consent may be required before any collection of data, and the users must be given access to their data, and/or the possibility to request its deletion. Actively initiated measurements, therefore, have advantages in terms of data privacy and security, since it is the user's conscious decision to provide the data or not.

Finally, one should be aware that business value is generated out of crowd data, which means that the data of each measurement have a certain value. Moreover, measurements have certain costs associated with them, even if it is just the power or data quota of the end user's device, or the end user's time. Thus, users which contribute crowd data should be reimbursed with an appropriate compensation for their data and their expenses.

## 3.2 Data Collection and Preparation

Having the design and methodology in place, and the data collected, efficient handling of the data becomes a challenge. One critical aspect of storing crowd data is their size and in some cases the method of containing it, for example, by using continuous streaming. Depending on the volume of the data, an appropriate data analytics framework should be used.

As described in the previous sections, the raw data may contain biased or unreliable records, which must be filtered out. The degree to which such data is invalid depends on the chosen methodology (e.g., subjective feedback is more prone to biases than automated measurements) and the desired representativeness. The role of the data scientist is to be able to answer questions about the origin and validity/reliability of the data, and – based on that – apply appropriate pre-processing steps.

Thus, the first step consists of understanding the data, which is often done by visualization. It includes to check expectations and distributions of given metrics, which may have to lie in a strict or expected range. However, even some out-of-range values may convey information. There might be a known issue/misbehaviour that is also a clear indication of an underlying problem. For example, a trace of GPS coordinates that includes invalid points may hint at a faulty or uncalibrated user device whose measurements must be excluded in general. Then, the general rule applies that the more that is known about the collected metrics, the easier the verification of the data validity becomes. Here, it is also helpful to cross-check different measurement metrics. For example, the collected radio technology can be used to filter out extremely good throughput samples, which could not be technically achievable with this radio technology. Finally, also data sources should be inspected with respect to validity, reliability, and representativeness. This not only applies when multiple crowd data sets are merged, but already for a single crowd measurement study. For example, data from users who are sending much more samples than others – especially for the user-initiated method – should be reviewed critically.

In general, depending on the confidence of understanding the data, the amount of cleaning can vary. For example, one may filter out only records based on statistical properties (e.g., apply 99% percentile filtering), or apply complex rules that require high confidence in the values (e.g., verify for subjective QoE ratings that a measured and rated video was actually watched). If a certain data cleaning rate (i.e., the percentage of records filtered out) is exceeded, root causes have to be investigated in order to be able to guarantee the general validity of the measurement approach, and if necessary, to recollect crowd measurements.

Pre-processing does not only mean cleaning the data. An important part of pre-processing is transforming and enriching the collected data with the help of models or other data sources.

First, one has to be aware of the layer on which the data was collected and analyses were performed, cf. Figure 3. Typically, network-layer measurements allow interpreting the Key Performance Indicators (KPIs) of networks or Quality of Service (QoS) metrics, such as throughput, packet loss, delay, or jitter. Application-layer measurements allow interpreting specific application behaviour, usually called application-layer QoS, or QoE-relevant metrics/KPIs, such as web page load times or video rebuffering. Measurements of user feedback allow interpreting the subjectively perceived quality with an Internet service, i.e., Quality of Experience (QoE). To transform the raw data from one layer into data on another layer, well-established and standardized models can be used. For example, throughput measurements could be translated with an appropriate model into page load times. Or, as shown in Figure 5, measurements of throughput can be used to estimate the behaviour of a video player, including its application-level KPIs like rebuffering and which video representation will be played at which time. This information can be used to estimate the QoE in terms of a Mean Opinion Score by use of an appropriate QoE model.

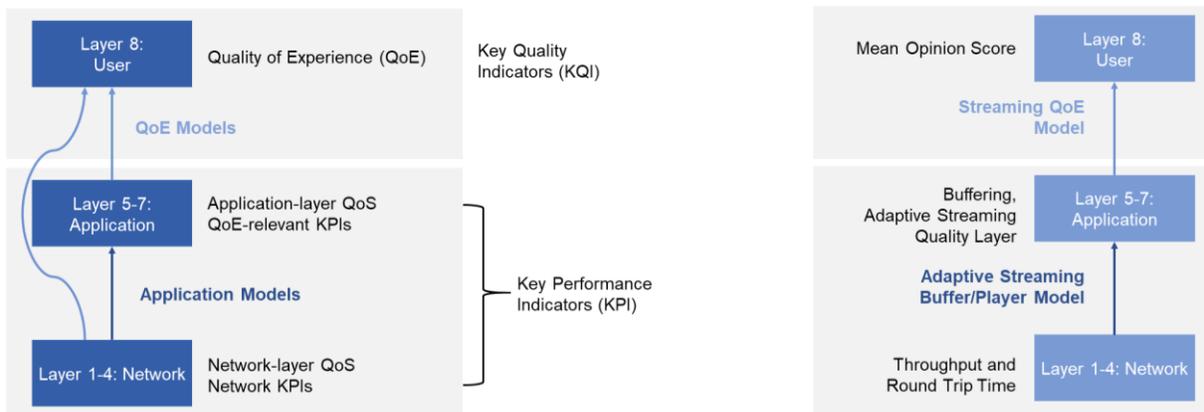

*Figure 5 Data Layers and Models including example for video streaming*

Second, metadata can complement crowd records in important ways. Metadata are a set of attributes about the measurements itself, e.g., the date of performing the measurement, or information about the device that collected the measurement. They allow to correctly use the measured metrics during analysis. Metadata can also be enhanced with other data sources. There are many context-related pieces of information that are not known by the reporting device, however – based on what is being reported – one can join that information with other data sources during pre-processing. For example, configuration details of mobile base stations can be added to the raw crowd data, which improves knowledge about the context and enables a greater potential for later analysis.

Finally, it is always good practice to store raw and pre-processed/cleaned data as separate data sets. Such an approach allows one to re-process all raw data based on new rules for pre-processing. This includes, for example, stricter filtering rules or new metrics.

## 3.3 Data Analysis

After the data have been collected and prepared, they can be analysed with respect to the research questions. Here, it is especially important to perform the analysis with respect to the specific decisions that have been taken in previous steps.

The first challenge in the analysis step is to reason about the representativeness of the data, as it defines to what extent the outcomes of the analyses can be generalized. This means that the analyst has to identify the specific population of the study, and to check if it corresponds to the intended population. Methodology provides the foundations for representativeness by selecting relevant participants and measurement conditions, nonetheless, the data might still be subject to sampling bias, such as systematic or sporadic measurement failures or external influences. For example, it could have happened that no representative set of participants could be recruited, or that the measurement did not execute properly on devices of some specific type. Moreover, it could have happened that measurements were influenced by external factors, such as limitations due to bandwidth or data caps. To detect sampling bias, it can be helpful to compare the measured data distributions to internal or external reference distributions, if available, such as demographic statistics or self-reports about end-users' network access speeds. Although it might be impossible to exclude all external influences, it is important to show awareness for sampling bias in order to apply statistical corrections or to modify the analyses and/or interpretations.

The second challenge concerns the validity of the analyses. To correctly analyse and interpret the data, it is a prerequisite to reason about the context of the measurement. The analyst has to clearly identify the independent and dependent variables of the study. At this stage, it could be helpful to evaluate the metadata of measurements, that is, descriptive data about the measurement process and its context, in order to be sure what is known and what can be inferred from the data, but also what is not known and what cannot be inferred from the data. Based on these reflections, appropriate metrics and statistical methods have to be selected. Typical pitfalls here include negligence of data types (e.g., computation of mean for ordinal data), insufficient aggregation of the sparse crowdsourced measurement data (e.g., too small sample sizes to detect significant differences), or unthoughtful merging of samples from different populations (e.g., merging tablet and smartphone measurements). The last point is especially important if the data stems from different data sources. Moreover, it also comprises the inverse situation, in which the data is inherently composed of different populations. Among others, multimodal distributions, or valid, but outlying measurements might be evidence of such situations. In this case, the different populations have to be identified and the data has to be partitioned accordingly before analysis.

The next challenge is the meaningful interpretation of the outcomes of the analyses. Especially in the context of crowdsourced network and QoE measurements, the analyst has to be aware of the layer on which the data was collected and analyses were performed. As described above, there are KPIs or QoS metrics from network-layer measurements (e.g., throughput), KQIs or QoE-related metrics from application-layer measurements (e.g., page load time), and there is the actual QoE based on subjective feedback. Being aware of these layers, it has to be carefully avoided to interpret on a different layer compared to the layer on which data was analysed. For example, it is not meaningful to infer web browsing QoE directly from the analysis of throughput measurements. Instead, there are well-established and standardized models, which can translate between the raw data on different layers, and thus, should be used. For the given example, the throughput measurements should first be translated with an appropriate model into page load times, then, the page load times should be translated with a web browsing QoE model (e.g., Egger, 2012) into QoE results. Here, it would also be possible to directly translate throughput measurements into QoE with the model in Casas (2016). There is a multitude of models proposed in the literature, thus, the analyst has to carefully select an appropriate model. Note that using a certain model might have certain requirements and might also limit the generalizability of the interpretations, thus, the applicability and the scope of the model have to be checked before the results are interpreted.

The final challenge, which is discussed here, focuses on ethical issues when evaluating and reporting the results of crowdsourced network and QoE measurements. As such studies include human participants, the analyst should comply with the ethical guidelines of its institution and with all legal regulations. This especially concerns privacy, such that participants cannot be deindividualized and no sensitive information is leaked. For this, it has to be ensured that analyses are always performed on aggregated datasets, which comprise measurement samples from larger groups of users, or that data is appropriately anonymized before further analyses, with only the variables available that are required for the defined goals of the analysis.

## 3.4 Summary

In the end, we want to note that this list of challenges cannot be considered exhaustive, but we think that it applies to most crowdsourced network and QoE measurement studies. Thus, awareness to the mentioned general challenges is highly recommended, as it will improve the value of the study outcome.